\documentclass[iop,apj]{emulateapj}
\usepackage{amsmath,amssymb,amstext}
\usepackage[breaklinks,colorlinks,citecolor=blue,linkcolor=magenta]{hyperref} 

\usepackage[all]{hypcap}
\def\baselinestretch{0.96}
\usepackage[normalem]{ulem}
\usepackage{graphicx}
\usepackage[flushleft]{threeparttable}
\usepackage{subfigure}
\usepackage{gensymb}
\usepackage{adjustbox}
\usepackage{array}
\newcolumntype{L}[1]{>{\raggedright\let\newline\\\arraybackslash\hspace{0pt}}m{#1}}
\newcolumntype{C}[1]{>{\centering\let\newline\\\arraybackslash\hspace{0pt}}m{#1}}
\newcolumntype{R}[1]{>{\raggedleft\let\newline\\\arraybackslash\hspace{0pt}}m{#1}}
\usepackage{array}
{\setlength{\extrarowheight}{5pt}
\usepackage{float}
\usepackage{natbib}
\begin{document}

\defcitealias{2018Natur.555...67B}{BRM18}
\defcitealias{2018Natur.564E..32H}{HKM18}

\title{The redshifted 21-cm signal in the EDGES low-band spectrum}
\author{ Saurabh Singh$^{1,2,3}$, Ravi Subrahmanyan$^{3}$}
\affil{{\small $^{1}$Department of Physics, McGill University, 3600 rue University, Montr\'eal, QC H3A 2T8, Canada}}
\affil{{\small $^{2}$McGill Space Institute, McGill University, 3550 rue University, Montr\'eal, QC H3A 2A7, Canada}}
\affil{{\small $^{3}$Raman Research Institute, C V Raman Avenue, Sadashivanagar, Bangalore 560080, India}}
\email{Email of corresponding author: saurabh.singh2@mcgill.ca}

\begin{abstract}

The EDGES collaboration reported the finding of an unexpectedly deep absorption in the radio background at 78~MHz and interpreted the dip as a first detection of redshifted 21-cm from Cosmic Dawn.  We have attempted an alternate analysis, adopting a maximally smooth function approach to model the foreground.  A joint fit to the spectrum using such a function together with a flattened absorption profile yields a best fit absorption amplitude of $921 \pm 35$~mK.  The depth of the 21-cm absorption inferred by the EDGES analysis required invoking non-standard cosmology or new physics or new sources at Cosmic Dawn and this tension with accepted models is compounded by our analysis that suggests absorption of greater depth.  Alternatively, the measured spectrum may be equally-well fit assuming that there exists a residual unmodeled systematic sinusoidal feature and we explore this possibility further by examining for any additional 21-cm signal.  The data then favors an absorption with Gaussian model parameters of amplitude $133 \pm 60$~mK, best width at half-power $9 \pm 3$~MHz and center frequency $72.5 \pm 0.8$~MHz.   We also examine the consistency of the measured spectrum with plausible redshifted 21-cm models:  a set of 3 of the 265 profiles in the global 21-cm atlas of \cite{2017MNRAS.472.1915C} are favored by the spectrum.  We conclude that the EDGES data may be consistent with standard cosmology and astrophysics, without invoking excess radio backgrounds or baryon-dark matter interactions.
 
\end{abstract}


\section{Introduction}

During the Dark Ages and subsequent Cosmic Dawn, differential cooling of matter against the cosmic microwave background (CMB) is expected to give rise to redshifted global 21-cm absorption features from neutral hydrogen at those times. The precise profile of the absorption is strongly tied to a number of astrophysical parameters \cite[]{2017MNRAS.464.1365M,2017MNRAS.472.1915C}.  The first ever detection of such an absorption feature was recently reported by the EDGES experiment \cite[][hereinafter BRM18]{2018Natur.555...67B}. The profile of the absorption feature was centered at 78~MHz and appeared flattened. The full width at half maximum was 19~MHz  and the depth of the absorption appeared to be 0.5~K. 

The reported detection was a complete surprise: both the amplitude and the flattening of the profile were unexpected and inconsistent with previous theoretical work based on astrophysical cosmology constrained by other data \cite[]{2017MNRAS.472.1915C}.   Therefore, the reported signal, if confirmed and genuine, has wide ranging implications for galaxy formation models \cite[]{2019MNRAS.483.1980M,2019arXiv190202438F}, dark matter models \cite[]{2018PhRvD..98f3021S,2018ApJ...859L..18S,2018JCAP...05..069M,2018PhRvD..98b3011L,2018arXiv180309739L} and also for expectations for the 21-cm power spectrum from Cosmic Dawn \cite[]{2018PhRvL.121a1101F,2018ApJ...864L..15K,2018PhRvL.121l1301M}.

The amplitude is a factor of two larger than the maximum amplitude considered possible based on standard theoretical predictions and cannot be explained by astrophysical models in the $\Lambda$CDM framework \cite[]{2018Natur.555...71B,2018PhRvD..97j3533W} or without finely tuned modifications to the background cosmology \cite[]{2018JCAP...08..037H}.  Physical explanations for the depth of the  absorption require invoking mechanisms that might cool the gas to temperatures below that achievable by adiabatic cooling \cite[]{2014PhRvD..90h3522T,2014PhRvD..89b3519D,2015PhRvD..92h3528M,2018Natur.555...71B}, or else by invoking an additional radio background at Cosmic Dawn apart from the CMB (\cite{2018PhRvD..98b3013S,2018ApJ...858L..17F}; \citetalias{2018Natur.555...67B}).

In recent literature motivated by the reported detection, the interaction between baryonic and dark matter has been considered as a new mechanism that might supplement adiabatic expansion and enhance the cooling of matter. Non-standard Coulomb-like interactions between dark-matter particles and baryons can lead to energy and momentum exchange between them and can cool the gas substantially \cite[]{2018Natur.555...71B, 2018PhRvD..98b3013S, 2018arXiv180210094M}. However, such interactions are severely constrained by other astronomical observations and laboratory experiments, which indicate that the dominant dark matter component is unlikely to be able to cool the gas to the extent indicated by the EDGES absorption \cite[]{2018arXiv180303091B,2018PhRvL.121a1102B,2018arXiv180210094M,2018PhRvD..98j3529K}. 

The second option of postulating an excess radio background in the high redshift universe at Cosmic Dawn has also been explored \cite[]{2018ApJ...858L..17F,2018ApJ...858L...9D}.  Accretion onto the first black holes has been suggested as a plausible source for an additional background. Such a scenario would also contribute to X-ray and UV backgrounds and thus require the sources to be sited in significantly obscured environments so that these radiations do not disallow the hyperfine populations to be consistent with the EDGES signal \cite[]{2018ApJ...868...63E}; additionally, their energetics are severely constrained by the substantial inverse-Compton losses at high redshifts \cite[]{2018MNRAS.481L...6S}.   Mechanisms involving particle interactions, often invoking exotic physics, have also been proposed to produce an excess radiation background that might explain the EDGES absorption \cite[]{2018arXiv180407340L,2018PhLB..783..301M,2018PhLB..784..130A}.

Detecting redshifted 21-cm from Cosmic Dawn (CD) and the Epoch of Reionization (EoR) is an extremely challenging experiment in long wavelength astronomy, requiring unprecedented measurement accuracy of the radiometer and control of systematics. Owing to the difficulty of such measurements, there have been concerns over the modeling of the EDGES data \cite[][hereinafter HKM18]{2018Natur.564E..32H}, suggestion that resonance modes in the ground beneath the antenna might lead to spurious absorption signatures \cite[]{2018arXiv181009015B}, and concerns regarding contamination from spinning dust \cite[]{2018ApJ...858L..10D}.  Motivated by the unexpected nature of the EDGES claim, and the difficulty with finding astrophysical explanations without introducing either new physics or new source populations as extensions to standard models for cosmology and particle physics, we have explored an alternate approach in the analysis of the EDGES spectrum.  

In this paper, we use publicly released EDGES data\footnote{\url{http://loco.lab.asu.edu/edges/edges-data-release/}}, and employ a different foreground model compared to that adopted by \citetalias{2018Natur.555...67B}.  Our approach is based on our earlier study \cite[]{2017ApJ...840...33S}  on modeling the foregrounds for such global CD/EoR measurements.  In Sec.~\ref{sec:1}, we briefly describe this maximally smooth function approach and present results from its application to the data.  In Sec.~\ref{sec:2}, we carry out joint modeling of the data using maximally smooth functions to represent the foreground together with flattened absorption with the form of the signal that is reported by EDGES.  We compare with the possibility that the calibrated data has residual systematics of sinusoidal form.   Adopting the hypothesis that there is such a residual in the data, in Sec.~\ref{sec:3} we perform joint fitting of a maximally smooth function to represent the foreground, plus sinusoidal systematic and examine for an additional 21-cm signal. Sec.~\ref{sec:dis} presents a discussion on the physical parameter space populated by favored signals.  We present a summary and conclusions in Sec.~\ref{sec:4}.

\section{Modeling the foreground in the EDGES spectrum as a maximally smooth polynomial}
\label{sec:1}

\begin{figure}[htbp]
\begin{center}
\includegraphics[scale=0.47]{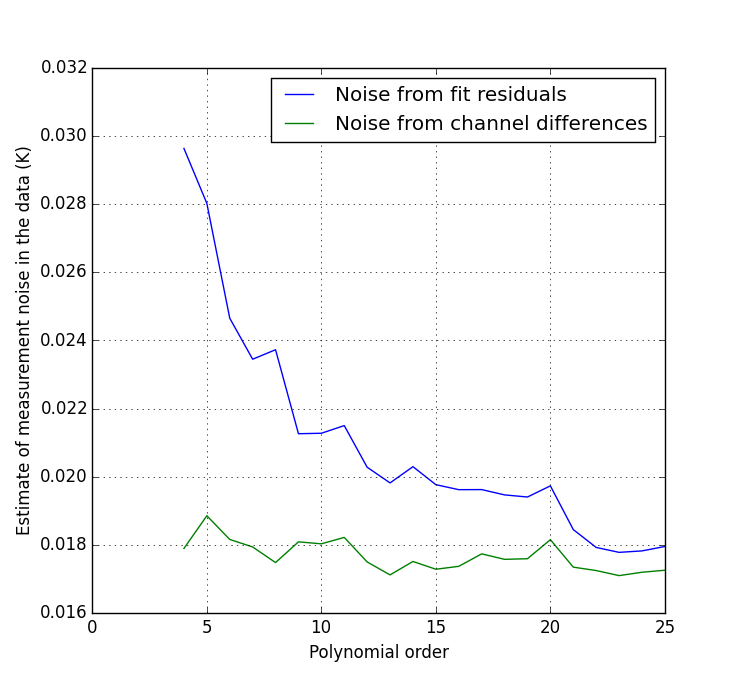}
\caption{Estimate of the measurement noise in the calibrated EDGES spectrum. The spectrum was fit, in log(temperature) versus log(frequency), with polynomials of increasing order.  The standard deviation in the residuals, corrected for the reduction in degrees of freedom for each polynomial order, is shown in the upper (blue) trace.  The lower (green) trace is the standard deviation of the residual spectrum estimated from the differences between adjacent channel data in the residuals, accounting again for the reduction in degrees of freedom for each polynomial order.}
\label{fig:0}    
\end{center}   
\end{figure}

We first estimate the instrument/measurement noise  in the EDGES spectrum.  The spectrum in log$_{10}$(temperature) versus  log$_{10}$(frequency) domain is fitted with polynomials of increasing order, the least-squares fit is subtracted from the data and the standard deviation in the residual is computed.  Additionally, for each residual spectrum, differences between every spectral value and the value in the adjacent spectral bin is computed, and the standard deviation in these differences is computed.  Since for a zero-mean Gaussian random  signal, this standard deviation in the differences is $\sqrt {2}$ times the standard deviation in the signal, we reduce these computed standard deviations by $\sqrt{2}$.  Fig.\ref{fig:0} shows the run of standard deviation in the residual on subtracting out best-fit polynomials of progressively greater order.   In computing these standard deviations, it may be noted that we have taken into account the progressive reduction in the number of degrees of freedom in the residual data when fitting with polynomials of progressively increasing order.  As shown by the figure, the standard deviation of residuals decreases with increasing order, indicating that the EDGES spectrum does have complex structure that requires a very high order polynomial to model to the sensitivity of the observations.  The standard deviations estimated from the differences between adjacent channels vary little with increasing order of the fit polynomial, as expected.  The analysis suggests that the instrument measurement noise in the spectrum corresponds to rms spectral fluctuations of about 17~mK.  

Radiometers, like EDGES, have wide beams that sample a large sky solid angle and hence their spectra represent the average over a large number of individual sources---compact and diffuse---which may have different emission mechanisms and spectral indices and shapes.  An average spectrum may no longer be accurately representable with a physical model that is appropriate for a single source.  For example, if the spectrum is the sum of spectra of two sources, and the pair have straight synchrotron spectra but with different spectral indices, the average will not fit a model of a synchrotron source with a single power-law spectrum.  In this case the spectrum at low frequencies would be dominated by the source with the steeper spectrum and at high frequencies the source with flat spectrum would dominate the total.  In log temperature versus log frequency domain, in which the individual sources have linear spectra, the total spectrum would have a curvature that would require higher and higher degree polynomial to fit with increasing accuracy and lower residuals.  For this reason, an attempt to use a physical model with multiple parameters to fit the total spectrum as measured by a radiometer may be limited in accuracy.

There have been a number of parametric and non-parametric approaches to foreground modeling \cite[]{2015aska.confE...5C,2009MNRAS.397.1138H,2015ApJ...799...90B}, most of which have no physical basis.  \citetalias{2018Natur.555...67B} have adopted a five-term parametric model to represent the foreground in the EDGES spectrum, the 5-term model is motivated by the physics of radiation mechanisms relevant to the frequency range of the spectrum, including synchrotron radiation with a curved spectral index, ionosphere emission and absorption.  \citetalias{2018Natur.564E..32H} point out that the best-fit parameters have unphysical values; as discussed in \citetalias{2018Natur.555...67B} and \cite{2018Natur.564E..35B}, there are residual instrumental effects and calibration errors that are expected to be subsumed by their 5-term foreground model and, therefore, the fit parameters may not be physical even if the foreground were completely described by their physical model.



\cite{2017ApJ...840...33S} created mock spectra that might be representative of sky spectra observed by radiometers with wide beams.  Using all-sky maps that are available at discrete frequencies, the spectra towards individual sky pixels were modeled by fitting a physical model (GMOSS; \cite{2017AJ....153...26S}) to the discrete measurements at that pixel.  The physical model for every pixel consisted of a synchrotron power law with a spectral break, or sum of two power laws, depending on whether the pixel spectrum was convex or concave respectively.  To this was added an optically-thin thermal emission component and the pixel spectrum was also allowed to have a low-frequency thermal absorption.  Mock spectra were generated, assuming observations with wide beams, by a beam-weighted averaging over the physical sky spectra of the individual pixels.  The work showed that the radiometer observations could be fit to mK accuracy using what was called ``maximally smooth'' functions, and that the residuals to such fits preserved the turning points inherent in any embedded 21-cm signals.  Because fitting with maximally smooth functions reveals in the residuals the turning points in any embedded signal, it can also be used as a diagnostic to inspect for non-smooth frequency structures in the data that may result from instrument generated additive signals or spectral distortions introduced in the intrinsic foreground spectrum by non-ideal, uncorrected properties of the antenna. 

As described in \cite{2015ApJ...810....3S,2017ApJ...840...33S}, maximally smooth functions are a special class of polynomials, where the second and higher order derivatives of the polynomial are not allowed to have zero crossings across the band of interest.  As a result, the constrained polynomial fits only to the smooth component of the spectrum and does not subsume any embedded wiggle or higher order variation present in the data.   If, for example, the foreground were a summation of synchrotron spectra with a distribution in spectral indices,  and had a 21-cm signal with multiple turning points as an additive, then fitting with a maximally smooth function would fit out the foreground and the residual would reveal the turning points.   As the order of the constrained polynomial-form maximally smooth function is increased, the smooth component of the spectrum is fitted with greater accuracy without also fitting out the 21-cm signal.  In contrast, fitting with unconstrained polynomials of increasing order would fit the foreground with increasing accuracy; however, the 21-cm signal would also be fitted out with greater accuracy and the residual would not obviously reveal the turning points.  

\begin{figure}[htbp]
\begin{center}
\includegraphics[scale=0.43]{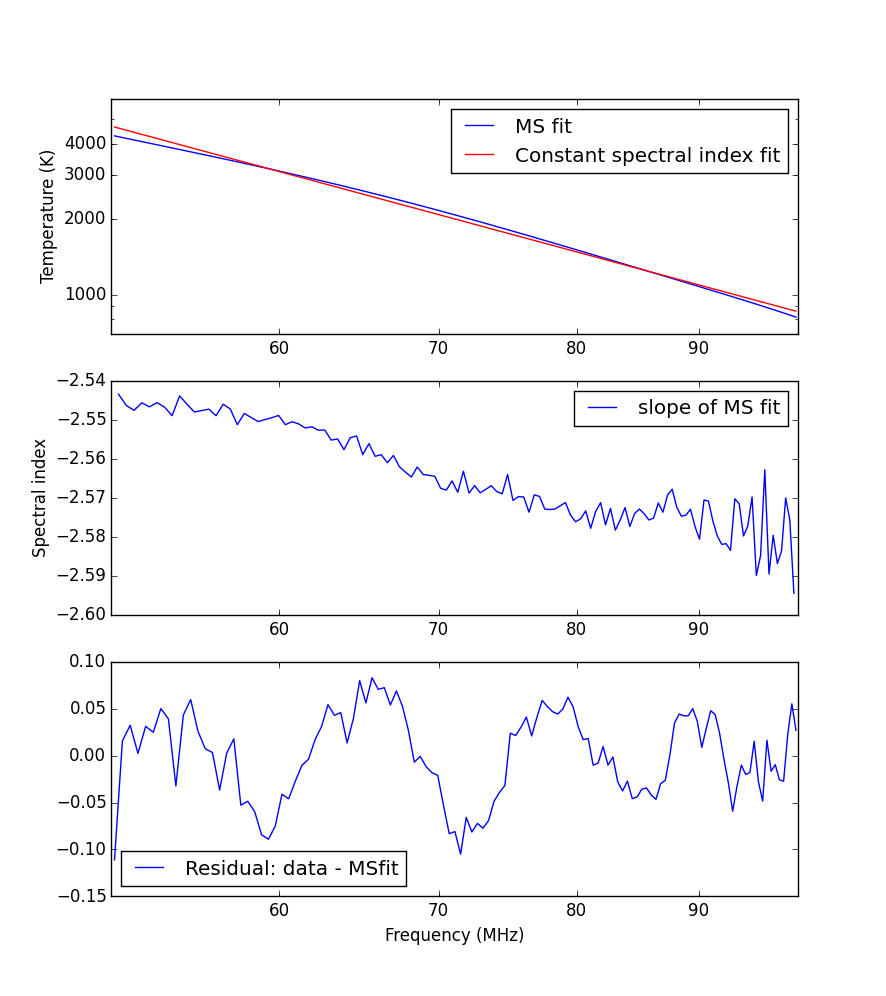}
\caption{The red straight line in the top panel is the best-fit to the EDGES data with a spectrum of constant spectral index.  The blue curved trace in the top panel is the best-fit maximally smooth (MS) function, with deviations from the straight-line fit amplified 30 times to better display the nature of the maximally smooth fit.  The middle panel shows the variation of spectral index across the band, as inferred from the maximally smooth fit.  The bottom panel shows the residuals after subtracting out the maximally smooth fit from the EDGES spectrum.}
\label{fig:1}    
\end{center}   
\end{figure}

Shown in the top panel of Fig.~\ref{fig:1} is the best fit maximally smooth function. The fit in Fig.~\ref{fig:1}, as well as the fits discussed below for different models, are least-squares fits assuming uniform errors across the spectrum. Ideally, the fits need to account for variation in errors across the spectrum that would arise from varying sky brightness and also systematic errors, which are difficult to estimate and may dominate the thermal noise from system temperature. For comparison and clarity in displaying the characteristics of this fit, the best fit power-law spectrum with constant spectral index is shown for reference in red and the best fit maximally smooth function is shown in blue with the deviations from the best-fit power law enhanced by factor 30.  Clearly, the EDGES spectrum is convex and steepens towards higher frequencies.  The run of spectral index is shown in the middle panel, where it is seen that the spectral index changes from about $-2.545$ at 50~MHz to about $-2.585$ at 100~MHz. It is evident from the middle panel of Fig.~\ref{fig:1} that the spectral index variation with frequency has an inflection point. The prominent inflection in the spectrum in log${_{10}}$(Frequency)-log$_{10}$(brightness temperature) domain is avoided by representing the spectrum in log${_{10}}$(Frequency)-brightness temperature; therefore, in this work we fit the EDGES spectrum with maximally smooth functions in the latter domain. It may also be noted here that throughout the analysis presented in this work, we use the EDGES spectrum over their full band of 50 to 100~MHz, which is the band adopted for analysis in \citetalias{2018Natur.555...67B}. 

We find that we require a maximally smooth function represented by a 6-th order constrained polynomial, with 7 polynomial terms, to model the smooth component in this spectrum to within the rms noise in the data. Unlike unconstrained polynomials, the effective number of degrees of freedom in the case of maximally smooth functions would almost always be less than the number of polynomial terms, provided that a sufficiently high order polynomial form is used. This is because of the constraints on higher-order derivatives, and hence on the coefficients, imposed by the condition that the polynomial be maximally smooth. We thus find that increasing the number of terms of the maximally smooth polynomial beyond 7 terms does not significantly improve the fit or change the best-fit parameters.  

The residual (data minus fit) is shown in the bottom panel of Fig.~\ref{fig:1} and has standard deviation 44~mK.  The standard deviation in the residual is significantly greater than the measurement noise in the data, which is about 17~mK, as discussed above.  The large residual clearly shows that the sky spectrum is inconsistent with expectations based exclusively on large-angle averages over physically motivated foreground models.   The residual to the maximally smooth fit shows that the EDGES data almost certainly does have an embedded component with multiple turning points, apart from the expected dominant foreground. 

\section{Spectral structures in EDGES data besides a maximally smooth foreground}
\label{sec:2}

We have demonstrated and discussed above that there is significant spectral structure in the EDGES data apart from a maximally smooth foreground.  We know of no algorithm that may invert the measured spectrum to decompose it into a maximally smooth foreground plus a unique additive signal.  The turning points suggested by the fit presented in the previous section provides a clue, which may serve to guess parametric forms for the additive embedded signal.  The data may then be jointly fitted with such parametric forms along with a maximally smooth polynomial that represents the foreground.  Any additive embedded signal, which when fitted to the spectrum jointly with a maximally smooth polynomial yields a residual that is consistent with thermal noise, is a plausible signal.  Below, we examine the residuals from fits that adopt signal profiles corresponding to the flattened absorption suggested by \citetalias{2018Natur.555...67B} and compare with residuals from adopting a sinusoidal function for the embedded signal.  

\begin{figure}[htbp]
\begin{center}
\includegraphics[scale=0.28]{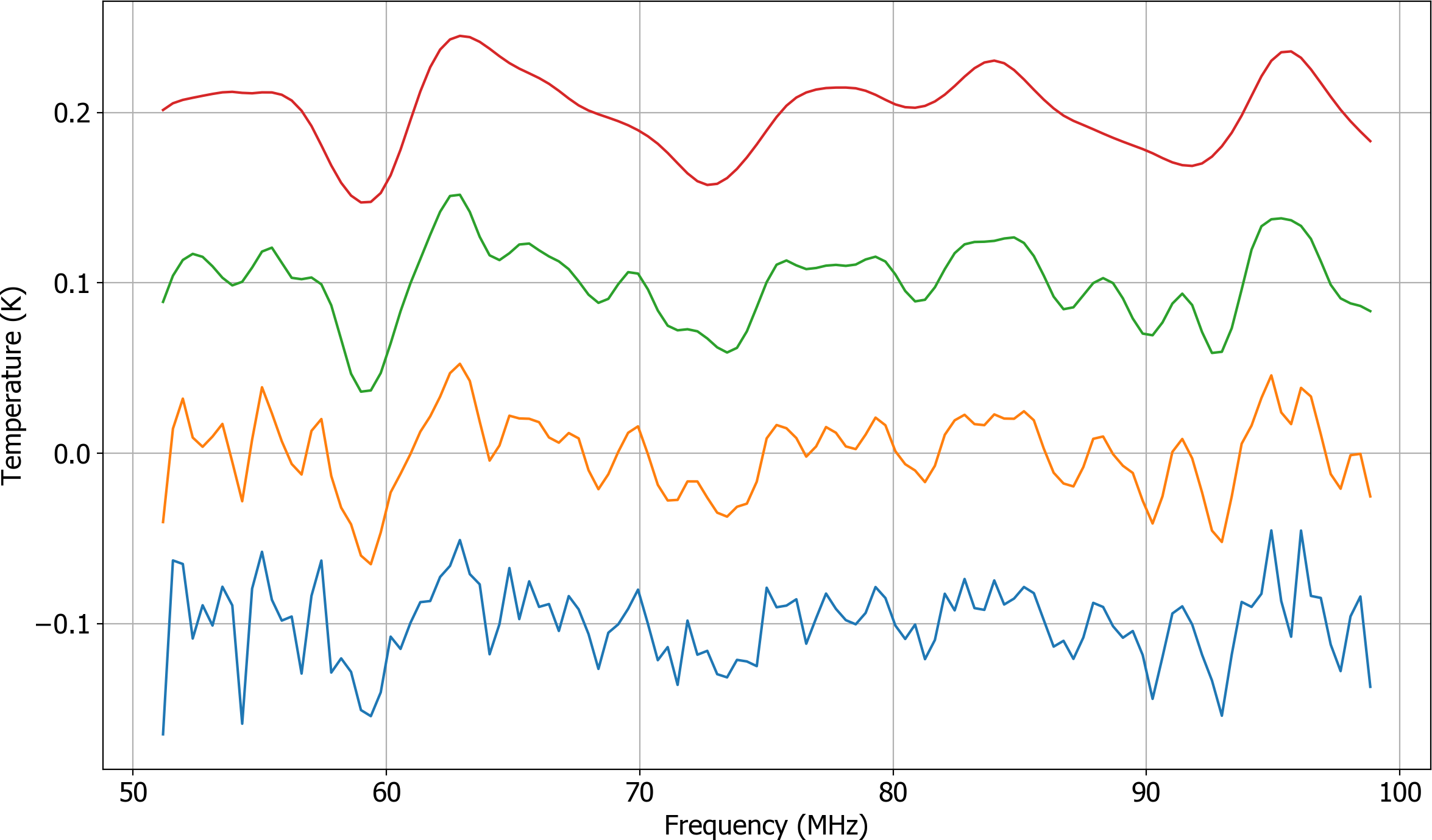}
\caption{The EDGES spectrum, with a flattened absorption profile with parameters suggested by \citetalias{2018Natur.555...67B} subtracted, was fit with a maximally smooth function.  The residual is shown in the lowest trace, offset vertically by $-0.1$~K.  Traces obtained by successively Hanning smoothing this residual with kernel widths of 0.781, 1.562 and 3.125~MHz are also shown, offset vertically by 0.1~K.  The smoothed traces have been individually scaled in amplitude to keep the standard deviations in the residuals to be the same as that in the unsmoothed residual, so that the structure in the smoothed residuals are clear in the display.  The standard deviation in the residual and in the three successively smoothed versions are, respectively, 23.1, 17.7, 14.2 and 10.5~mK.}
\label{fig:BDMS}    
\end{center}   
\end{figure}

\begin{figure}[htbp]
\begin{center}
\includegraphics[scale=0.28]{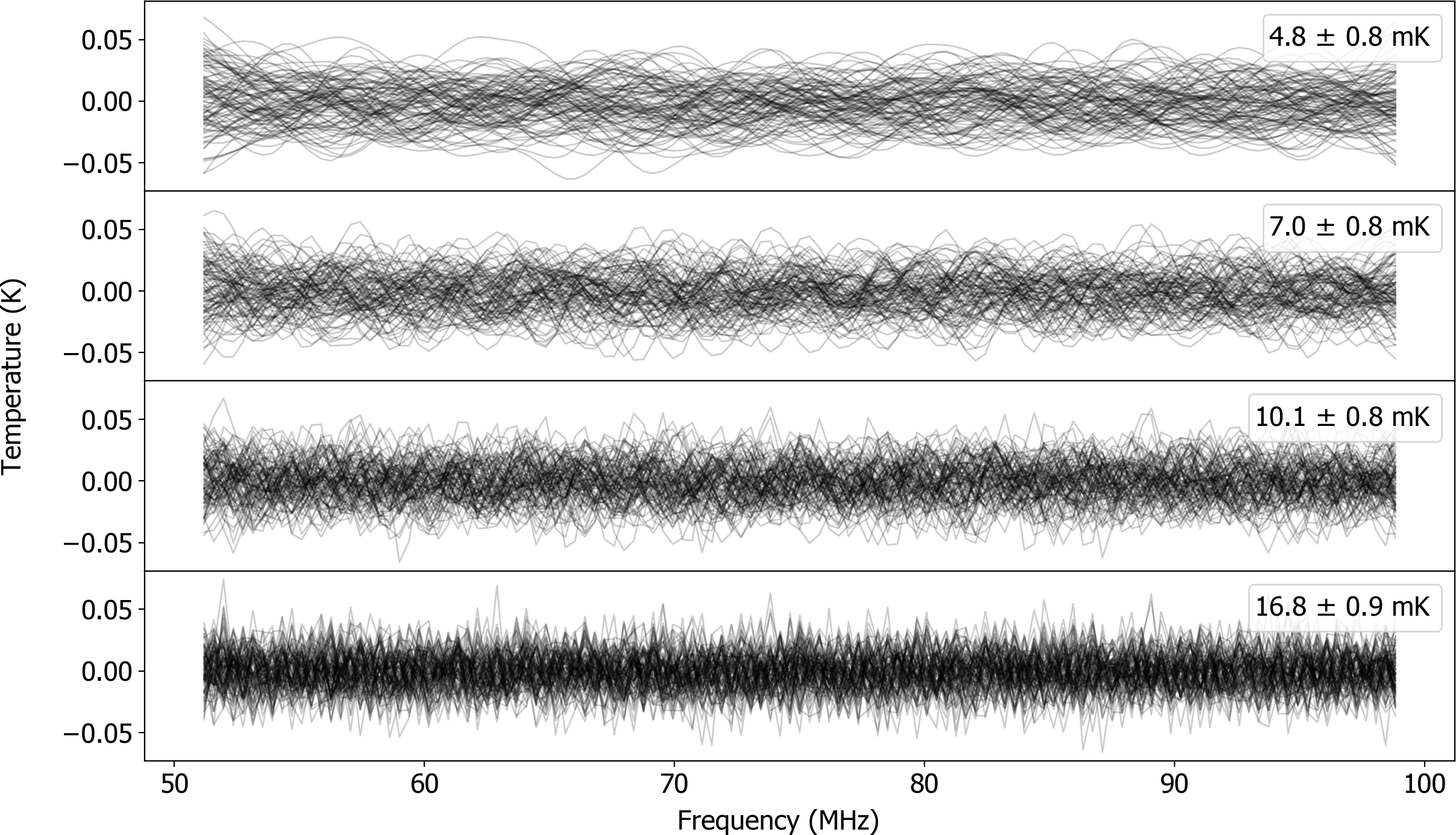}
\caption{As an example of what is expected of the residuals if the model represented the measurement well, we show in the lowest trace an example of 100 Gaussian noise realizations with the variance expected of the residuals. All the realizations are drawn from a Gaussian distribution of standard deviation 17~mK.  The traces above that are Hanning smoothed versions of this residual. The scaling method for smoothed versions is same as that in Fig.~\ref{fig:BDMS}.  The mean of the standard deviations in the Gaussian noise spectra in the lowest set of traces, and in the three successively smoothed versions above that, are 16.8, 10.1, 7.0 and 4.8~mK respectively. The uncertainty in these estimates, 0.8-0.9~mK, is computed from the spread of standard deviation in different noise realizations. These provide a reference for comparing with residuals and smoothed residuals obtained for different models considered herein.}
\label{fig:12}    
\end{center}   
\end{figure}

We begin this section with first assuming that the data has a flattened absorption profile with the form 
\begin{equation}
T_{21}(\nu) = -A \biggl( {1 - e^{-\tau e^{B}}\over{1-e^{-\tau}} } \biggr),
\end{equation}
with
\begin{equation}
B = { {4(\nu - \nu_0)^2}\over{w^2} }    {\rm log} \biggl[ -{1\over{\tau}} {\rm log} \biggl( { {1+e^{-\tau}}\over{2} } \biggr) \biggr],
\end{equation}
and parameters suggested by \citetalias{2018Natur.555...67B}: $A$ = 0.50, $\nu_0$ = 78.0, $w$ = 19.0 and $\tau$ = 7.0.  Having first subtracted this profile from the EDGES spectrum, we then fit the remainder with a maximally smooth function.   The final residual after subtracting both the flattened absorption with the \citetalias{2018Natur.555...67B} parameters plus this best fit maximally smooth function is shown in Fig.~\ref{fig:BDMS}.  The residual has standard deviation 23.1~mK.  

Smoothing the residuals enhances signal-to-noise for broader-scale frequency structures that are coherent across frequency channels \cite[]{Press:2007:NRE:1403886}. Therefore, the residual, Hanning smoothed \cite[]{oppenheim1999discrete} to successively lower frequency resolutions, is also shown in Fig.~\ref{fig:BDMS}.  For comparison and as a reference, we show in Fig.~\ref{fig:12} sample residuals containing thermal noise realizations drawn from a Gaussian distribution with zero mean and 17~mK standard deviation, along with smoothed versions of this synthetic spectrum.  The standard deviation in this case reduces to 4.8~mK at the maximum smoothing scale of 3.125~MHz.  Comparing the residuals in Fig.~\ref{fig:BDMS} with the expectation in Fig.\ref{fig:0}, there is clearly unmodeled structure present in the residual and the standard deviation does not attain that expected from measurement noise. Additionally, comparing the smoothed versions of the residual in Fig.~\ref{fig:BDMS} with the reference set in Fig.~\ref{fig:12}, it is clear that smoothing the residual does not decrease the standard deviation of the smoothed spectrum as expected given the width of the smoothing functions: Hanning smoothed with kernel of width 3.125~MHz, the standard deviation in the residual reduces to only 10.5~mK.

\begin{figure}[htbp]
\begin{center}
\includegraphics[scale=0.28]{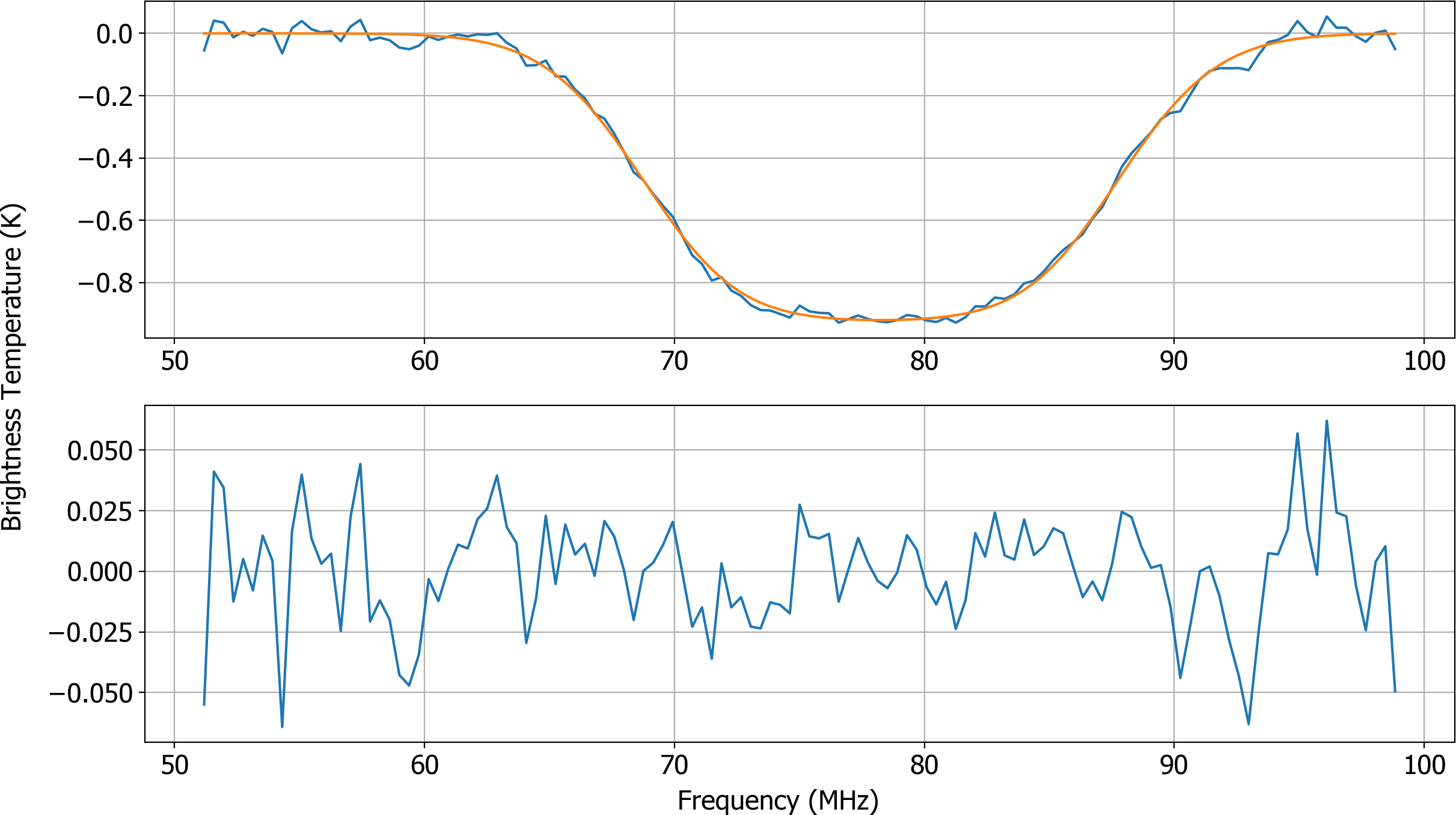}
\caption{The result of jointly fitting the EDGES spectrum with a maximally smooth function plus flattened absorption profile.  The top panel shows the residuals after subtracting out only the maximally smooth fit from the measured spectrum, with the best-fit flattened absorption profile overlaid. The bottom panel shows the residuals after both components of the joint fit are subtracted from the measured spectrum.}
\label{fig:4}    
\end{center}   
\end{figure}

\begin{figure}[htbp]
\begin{center}
\includegraphics[scale=0.28]{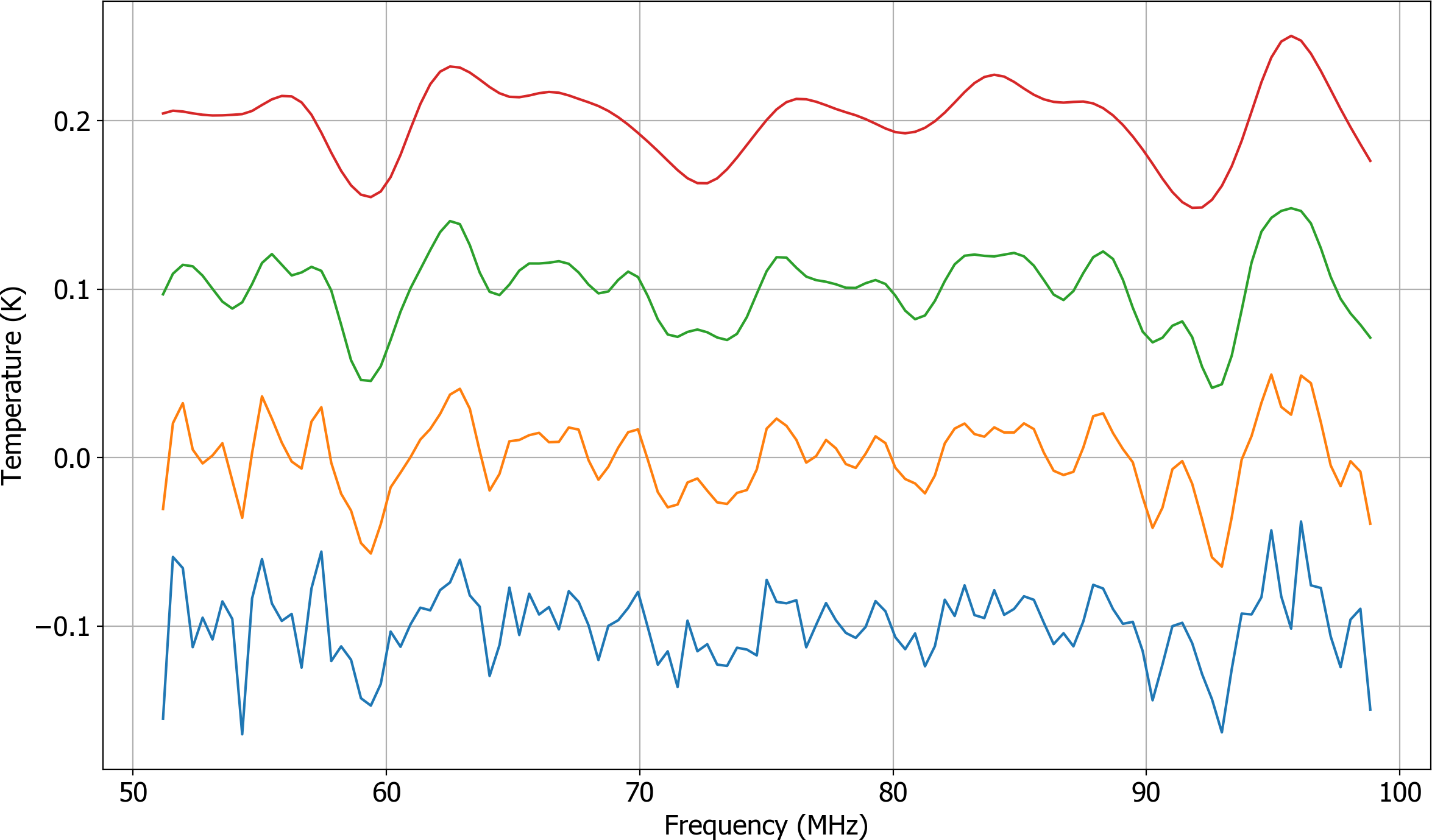}
\caption{Residual of the joint fit with a maximally smooth function plus flattened absorption profile is shown in the bottom trace.  Also shown in the traces above that are Hanning smoothed versions.  The offsets and scaling are as in Fig.~\ref{fig:BDMS}.  The standard deviation in the residual and in the three successively smoothed versions are, respectively, 22.5, 16.9, 13.2 and 9.0~mK.}
\label{fig:4b}    
\end{center}   
\end{figure}

We next perform a joint fit, of a maximally smooth function and the flattened absorption profile, this time allowing the parameters of the absorption to be optimized. The flattened profile is described by its amplitude $A$, centre frequency $\nu_0$, full width at half-power $w$ and flattening factor $\tau$, and we keep all the four parameters unconstrained in the modeling. The outcome of the fit is shown in Fig.~\ref{fig:4} and the residuals, smoothed to successively lower spectral resolutions, is shown in ~\ref{fig:4b}. The best fit amplitude is $921 \pm 35$~mK centered at frequency $78.2 \pm 0.1$~MHz. Best-fit full width at half maximum is $19.2 \pm 0.2$~MHz and the flattening factor is $4 \pm 0.28$.  The resulting standard deviation in the residual structures is 22.5~mK, which reduces to 9.0~mK when Hanning smoothed with kernel of 3.125~MHz. It may be noted here that the best fit depth of the absorption we have obtained differs significantly from that reported by \citetalias{2018Natur.555...67B}, lying at the boundary of their $99\%$ confidence interval.  If the foreground were modeled as a maximally smooth function instead of the 5-term physical model proposed by \citetalias{2018Natur.555...67B}, the absorption is inferred to have an amplitude of about 0.9~K instead of the 0.5~K inferred by \citetalias{2018Natur.555...67B}.  As we have argued above, the foreground component in wide-beam radiometer measurements are not expected to follow physically motivated forms appropriate for individual sources, but are expected to be fit with maximally smooth functions.  Therefore, if the 21-cm signal from Cosmic Dawn has the form of a flattened absorption profile, the depth of the absorption is 0.9~K, and the tension with previous theoretical expectations is all the more worsened.

\begin{figure}[htbp]
\begin{center}
\includegraphics[scale=0.28]{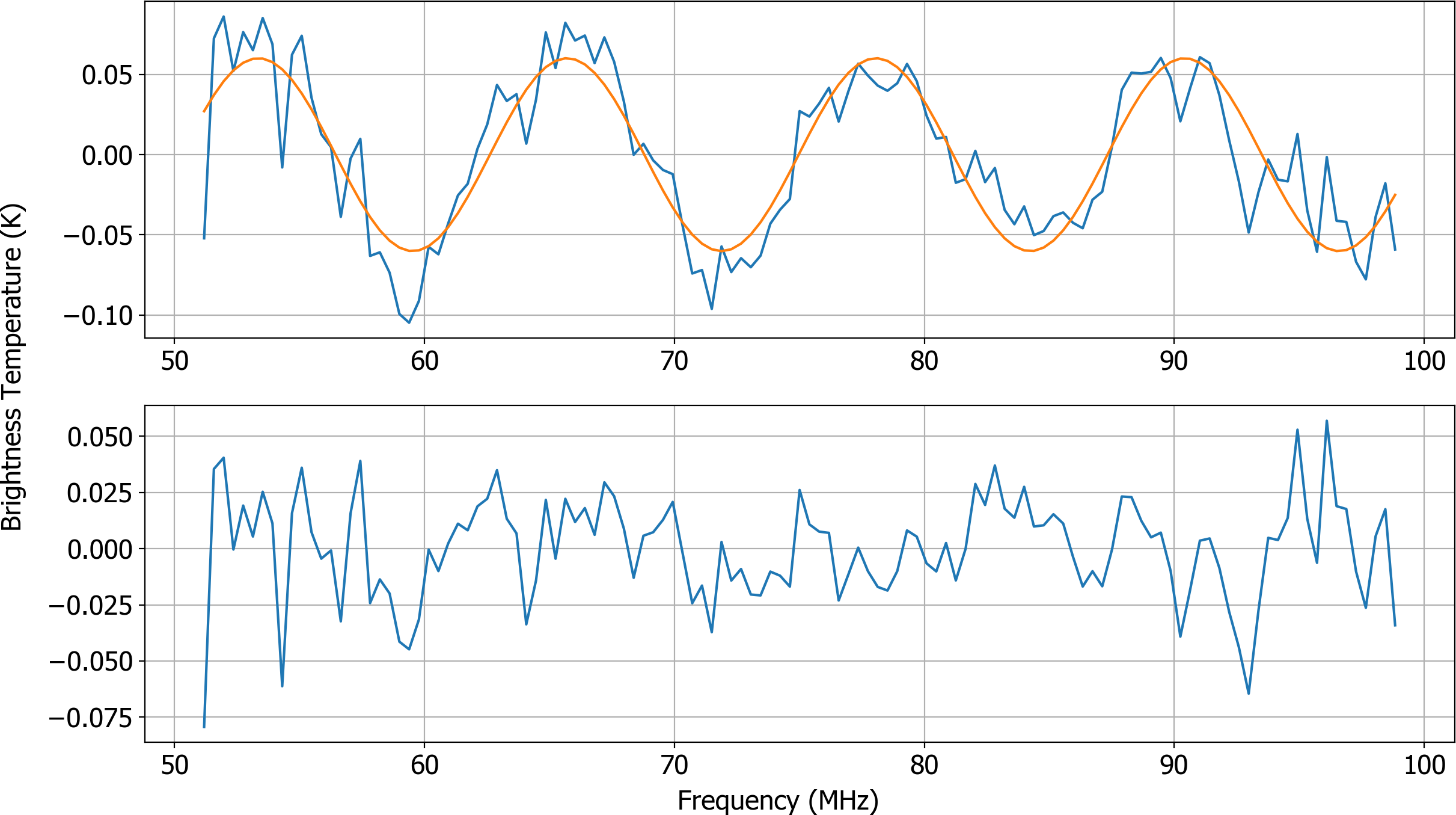}
\caption{The result of jointly fitting the EDGES spectrum with a maximally smooth function plus sinusoid. The top panel shows the residuals after subtracting out only the maximally smooth component of this joint fit from the measured spectrum, with best-fit sinusoid overlaid. The bottom panel shows the residuals after both components of the joint fit---maximally smooth component and the sinusoid---are subtracted from the measured spectrum.}
\label{fig:3}    
\end{center}   
\end{figure}

\begin{figure}[htbp]
\begin{center}
\includegraphics[scale=0.28]{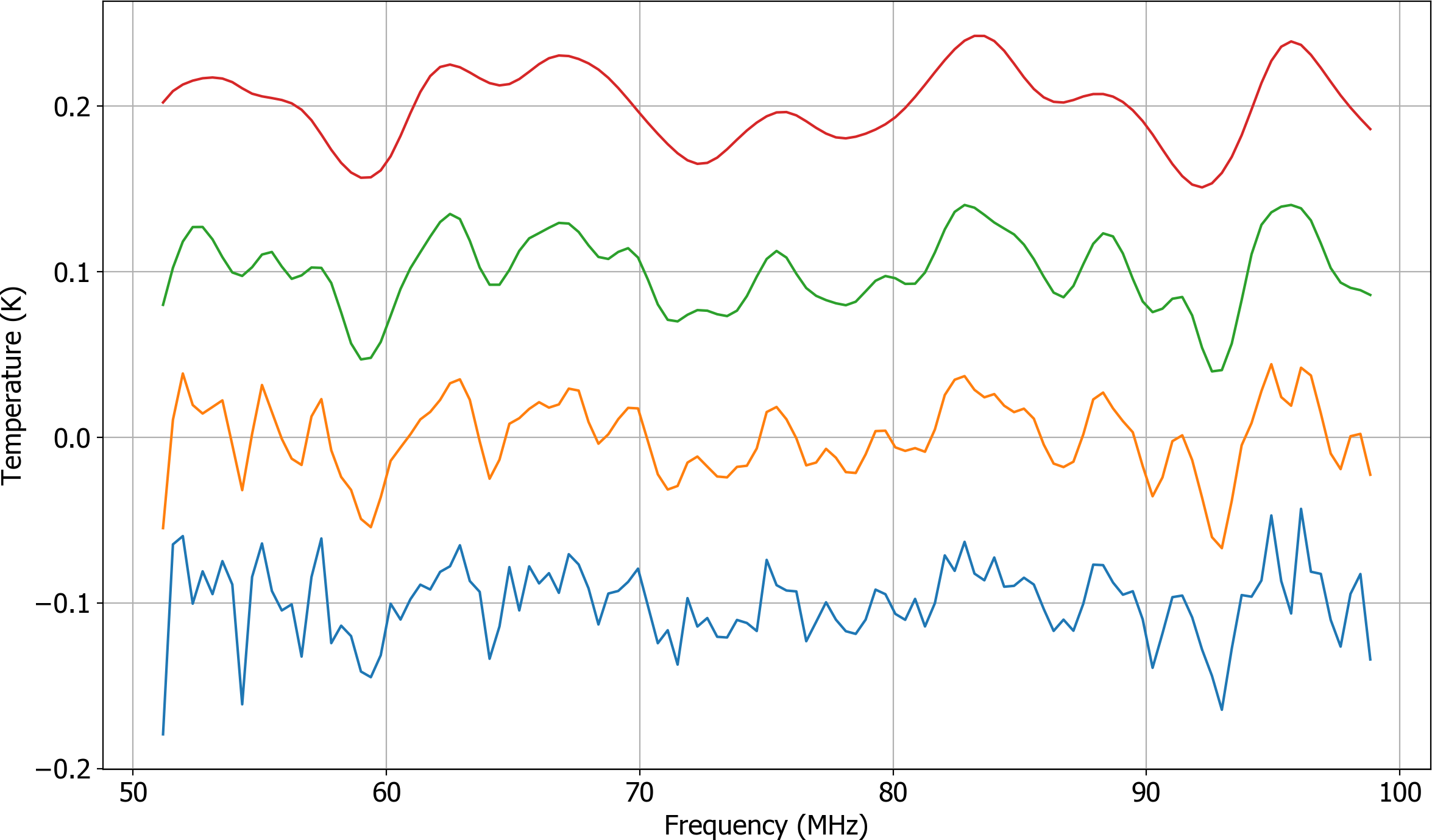}
\caption{The residuals of the joint fit with maximally smooth function plus sinusoid is shown in the bottom trace with mean offset by $-0.1$~K.  The offsets and scaling are as in Fig.~\ref{fig:BDMS}.  The standard deviation in the residual and in the three successively smoothed versions are, respectively, 22.9, 17.2, 13.3 and 9.4~mK.}
\label{fig:3b}    
\end{center}   
\end{figure}

Motivated by the above finding, and also the periodic sinusoidal nature of the residual structure in Fig.~\ref{fig:1}, we have attempted to model the spectrum jointly with a maximally smooth function and a sinusoid with amplitude, frequency and phase as free parameters.  In Fig.~\ref{fig:3}, we show the best fit results and in Fig.~\ref{fig:3b} we shown the residuals Hanning smoothed to lower resolutions.  The best fit amplitude of the sine function is $60 \pm 10$ mK, with a period of $12.3 \pm 0.1$ MHz. This is similar to the result of \citetalias{2018Natur.564E..32H} where the foreground model was the same as that adopted by \citetalias{2018Natur.555...67B}.  The standard deviation in the residuals after subtracting the best-fit maximally smooth foreground and best fit sinusoid is 22.9~mK, employing three free parameters to model the non-smooth component. The resulting standard deviation is only marginally higher than that obtained assuming a flattened absorption profile for the non-smooth component, which uses four free parameters.  Smoothed with a Hanning kernel of width 3.125~MHz, the residuals in the case of assuming a sinusoidal form for the signal have standard deviation of 9.4~mK, which is again marginally higher than that for the case of assuming a flattened Gaussian form signal.  

Although the residuals in the case of assuming a sinusoidal form for the signal has slightly higher standard deviation, sinusoidal signals have smaller number of free parameters compared to that for a flattened-absorption type signal.   Therefore, to compare the goodness of fits and usefulness of the modeling in the two cases, we compute the Bayesian information criterion (BIC) for the two cases.

Assuming normal distribution of model errors, BIC can be defined as
\begin{equation}
\textrm{BIC} = n\,\textrm{ln}\left( \frac{1}{n}\sum_{i=1}^{n}(x_i - \hat{x_i})^2 \right) + k\,\textrm{ln}(n),
\end{equation}
where $n$ is the number of data samples, $k$ is the number of parameters and $\sum_{i=1}^{n}(x_i - \hat{x_i})^2$ is the reduced sum of squares \cite[]{1978AnSta...6..461S,9780125649223}.  The model with lower BIC is preferred.  We take the number of free parameters for the maximally smooth function to be the number of terms in its polynomial expansion, though the effective number of free parameters would be lesser as discussed above in Sec.~\ref{sec:1}. We thus present the worst-case scenario for BIC by making this assumption. Since we use the full band dataset, the number of data samples remain the same, $n=123$, in all cases. We adopt the scale in ~\cite{doi:10.1080/01621459.1995.10476572} for interpreting BIC values, in which it is argued that if the difference in BIC values for two models, $\Delta \textrm{BIC}$, exceeds 6, then there is  strong evidence against the model with the higher BIC. Unsurprisingly, both the sinusoidal and flattened absorption profiles result in very similar BIC values of $-876.3$ and $-876.1$ respectively ($\Delta \textrm{BIC}\sim 0.2$), which confirms the finding that neither of these models ought to be preferred over the other, at least in so far as this formal statistical comparison is concerned \cite[]{doi:10.1080/01621459.1995.10476572}.  It is therefore equally likely that the data may well have embedded a sinusoidal unwanted systematic of hitherto unknown origin, instead of an unexpectedly deep flattened 21-cm absorption. 

Sinusoidal systematics may arise due to impedance mismatch between the antenna and subsequent electronics in the receiver, or along the analog signal path. As a result, all receiver noise in the signal path within the receiver electronics will suffer multipath propagation owing to reflection at the antenna and also at impedance mismatches in the signal path.  Multipath propagation within the signal path leads to sinusoidal spectral structure in the measured spectrum \cite[]{meys1978wave}. If the foreground sky signal suffers multipath propagation, the resulting sinusoid amplitude would directly scale with the sky brightness temperature. However, when receiver noise suffers multipath propagation, the amplitude of the sinusoid would be constant in time to the extent that the receiver noise is constant. Since \cite{2018Natur.564E..35B} do not observe a change in the amplitude of the flattened Gaussian model with change in sky brightness temperature, we infer that the sinusoid systematic might be an additive component arising due to multipath propagation of receiver noise.

It may be noted here that the best fit period of the sinusoid is close to the spectral periodicity expected in the antenna gains, as shown in Fig.~4 (b) and (c) in \citetalias{2018Natur.555...67B}. The period for the sinusoid that we infer to be likely present as a spurious component in the EDGES low-band data is close to half of the period of the spectral ripple that appears present in the residuals of the published EDGES high-band results (see Fig.~4(b) in \cite{2017ApJ...847...64M}).  The amplitude of the best-fit sinusoid in the low-band case is about 60~mK, and this is very similar to the amplitude of the sinusoidal feature in the residual spectrum of \cite{2017ApJ...847...64M}. The low-band antenna is a factor of two scaled-up version of the high-band antenna. Such a scaling would have the same impedance properties in corresponding bands, and hence the amplitude of sinusoid is expected to be the same in the two bands. Given that this is consistent with what we observe, it is possible that the sinusoidal ripple is an inherent characteristic related to the antenna structure or its interaction with the environment.

In summary, (i) on fitting with just a maximally smooth function (Fig.~\ref{fig:1}) the residual was very suggestive of an additional sinusoidal component, (ii) multipath propagation and reflections at impedance mismatches within the signal path both lead to sinusoidal spurious,  (iii) interpreting the EDGES data as implying an expectedly deep absorption causes substantial tension with prevalent theoretical models based on previous observations and (iv) Bayesian information criterion computed in the two cases, taking into account the standard deviations in the residuals in the two cases along with the numbers of free parameters in describing the non-smooth flattened absorption profile and sinusoidal profile, suggest that neither of these models are preferred over the other.  For all these reasons taken together, we model the data as containing a foreground of maximally smooth functional form plus a sinusoidal spurious component, and test in Sec.~\ref{sec:3} below for the presence of any additional cosmological 21-cm signals.

\section{The 21-cm signal from Cosmic Dawn in the EDGES data}
\label{sec:3}

\begin{figure}[htbp]
\begin{center}
\includegraphics[scale=0.28]{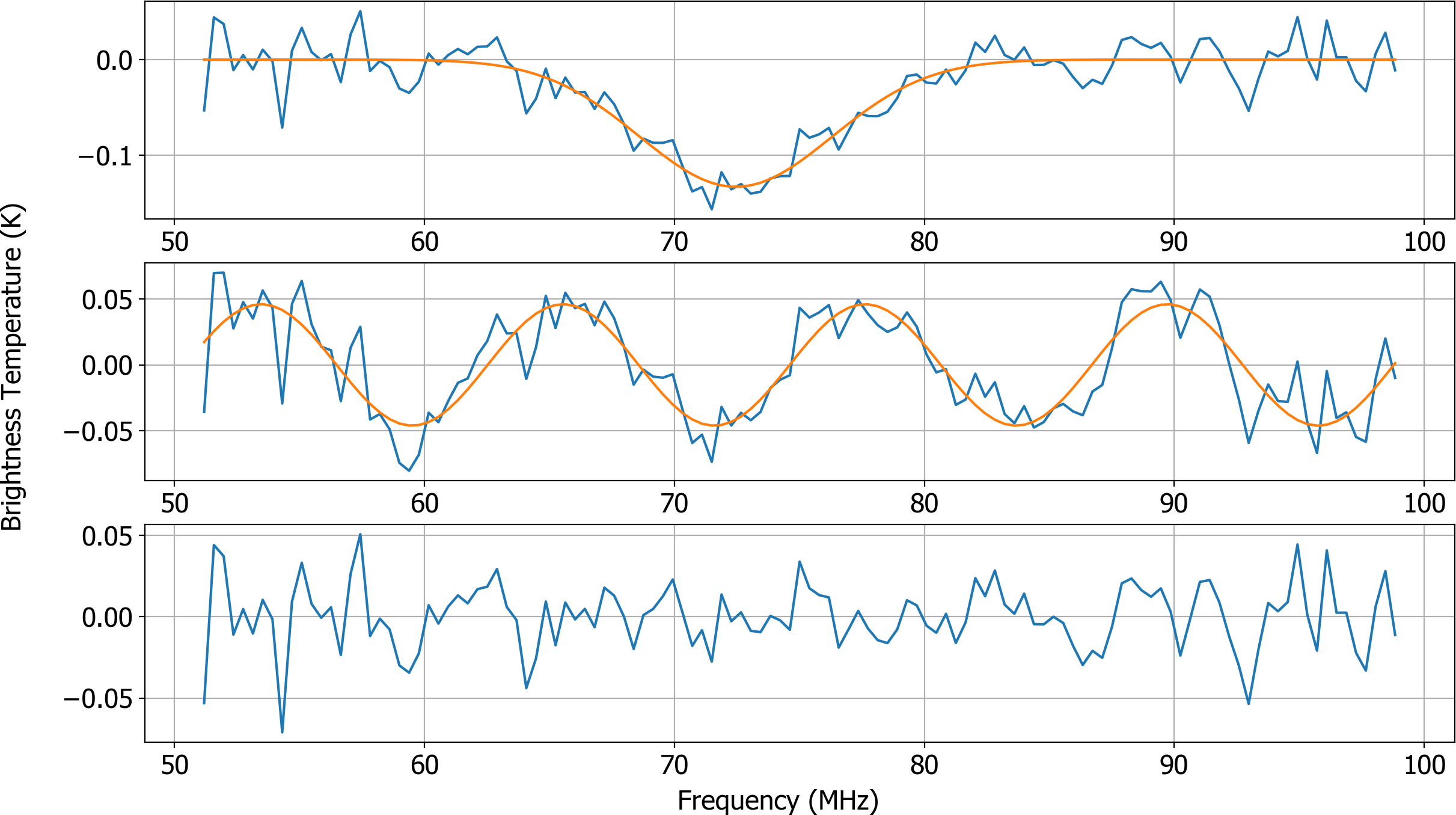}
\caption{Result of a joint fit to the EDGES spectrum with a maximally smooth function, plus a sinusoid, plus a Gaussian.  The top panel shows the spectrum, after subtracting the maximally smooth function and sinusoid, with the best-fit Gaussian overlaid. The middle panel shows the spectrum, after subtracting the maximally smooth function and Gaussian, with the best-fit sinusoid overlaid. The bottom panel shows the residuals after subtracting all three components in the best fit model. The standard deviation of the residuals is 20.1~mK, which is a significant reduction from that of the residual obtained by \citetalias{2018Natur.555...67B} after their subtraction of their best-fit foreground and 0.5~K flattened absorption signal.}
\label{fig:6}    
\end{center}   
\end{figure}

\begin{figure}[htbp]
\begin{center}
\includegraphics[scale=0.28]{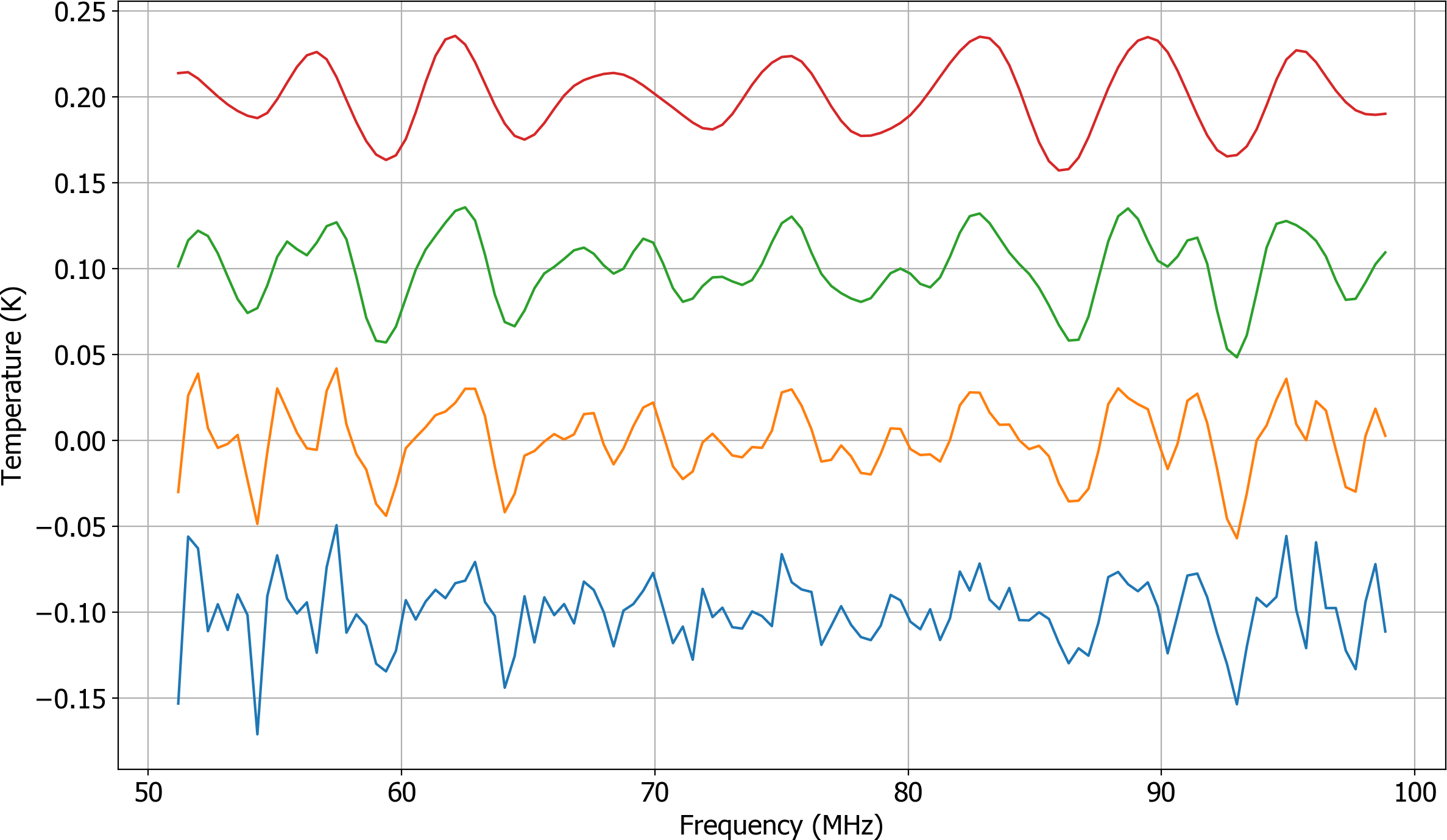}
\caption{Residual of the joint fit with maximally smooth function plus sinusoid plus Gaussian is shown in the bottom trace with mean offset by $-0.1$~K.  The offsets and scaling are as in Fig.~\ref{fig:BDMS}.  The standard deviation in the residual and in the three successively smoothed versions are, respectively, 20.1, 13.9, 9.4 and 4.6~mK.}
\label{fig:6b}    
\end{center}   
\end{figure}

As discussed above, we adopt a combination of a maximally smooth function along with a three-parameter sinusoid to represent, respectively, the foregrounds and residual systematic structure in the EDGES spectrum. With this model, we inspect the data for the presence of any additional physically motivated 21-cm signals. Since most accepted theoretical models predict an absorption feature in Cosmic Dawn as a result of first a Wouthuysen-Field coupling of the spin temperature to the low kinetic temperature, followed by X-ray heating, we examine for an additional Gaussian form absorption signal.   We parameterize the absorption signal to be of Gaussian form with amplitude, frequency and width as free parameters \cite[]{2016MNRAS.461.2847B}.  A joint fit of a maximally smooth function, plus a sinusoid, plus a Gaussian-form absorption yields a best-fit amplitude of $133\pm 60$~mK for the Gaussian, with width at half-power of $9 \pm 3$~MHz and centered at $72.5 \pm 0.8$~MHz.  This best fit model is shown in Fig.~\ref{fig:6}, with smoothed versions of the residuals shown in Fig.~\ref{fig:6b}.  We find that including the Gaussian appreciably reduces further the standard deviation of the residuals to 20.1~mK; smoothing the residuals with Hanning kernel of width 3.125~MHz yields a residual with standard deviation 4.6~mK.   In comparison, the joint fit of the 5-term physical model describing the foreground together with a 4-term flattened absorption profile, adopted by \citetalias{2018Natur.555...67B}, results in residual with standard deviation of 24.6~mK. Though the number of free parameters in the model adopted here is larger, information metrics like BIC that take into consideration the reduction in standard deviation along with the number of free parameters strongly favor the model suggested here, which includes a sine-form spurious additive, over the model adopted in \citetalias{2018Natur.555...67B} ($\Delta \textrm{BIC}\sim 26$). Comparing different models for the spectrum based on BIC,  model adopted in \citetalias{2018Natur.555...67B} is least preferred, followed by a maximally smooth function plus a flattened Gaussian profile ($\Delta \textrm{BIC} \sim 8$). The most preferred model is a combination of maximally smooth function, sinusoid and Gaussian: when compared to maximally smooth function plus a flattened Gaussian profile, this model is preferred by the Bayesian information metric with a $\Delta \textrm{BIC} \sim 18$. Table~\ref{tab:2} summarizes the models explored, along with the standard deviation of the residuals and BIC.

\begin{table*}[htbp]
{
\centering
\caption{Summary of different models explored using the EDGES measured spectrum.}
\label{tab:2}
\begin{tabular}{||cccc||}
\hline
Model & Number of free parameters & Residual rms (mK) & BIC \\
\hline
\hline
5-term foreground  + Flattened Gaussian (adopted in \citetalias{2018Natur.555...67B}) & 9 & 24.6 & -867.7 \\
\hline
MS + Flattened Gaussian & 12 & 22.5 & -876.1 \\
\hline
MS + Sinusoid & 11 & 22.9 & -876.3 \\
\hline
MS + Sinusoid + Gaussian & 14 & 20.1 & -894.2 \\
\hline
\end{tabular}
}
\begin{tablenotes}
     \small
      \item MS denotes a maximally smooth function. For uniformity, we use an 8-term maximally smooth function for all the models listed here, and assume that the number of parameters for the maximally smooth functions is the same as the number of terms. Given the constrained nature of maximally smooth functions, the number of free parameters in the case of maximally smooth functions will be less than the number of terms; therefore, this assumption is conservative. $\Delta \textrm{BIC} \geq 6$ suggests strong evidence against the model with the higher BIC \cite[]{doi:10.1080/01621459.1995.10476572}; which in our case would be the model that is less negative.
    \end{tablenotes}
\end{table*}

\cite{2017MNRAS.472.1915C} have provided an atlas of 265 theoretically plausible global 21-cm signal templates that were generated by semi-numerical simulations consistent with constraints from previous observations that constrain the physics and sources through Cosmic Dawn and reionization.  The atlas sparsely covers the signal space, and spans the range of absorption profiles expected in Cosmic Dawn.  We have examined whether these templates might be allowed by the EDGES data, apart from a maximally smooth foreground and sinusoidal spurious.  The EDGES spectrum is examined for each of the templates in the atlas, one by one, by performing a joint fit of the EDGES data to a maximally smooth function, sinusoid, plus the 21-cm signal template $S_0(\nu)$ multiplied by a scale factor $c$, which is a free parameter in the joint fit. The order of the maximally smooth function is chosen large enough so that the resulting residuals do not change with further increase in the number of terms of the polynomial. This approach is similar to the method adopted in \cite{2017ApJ...847...64M} and \cite{0004-637X-858-1-54}.  We select those 21-cm signal templates of the atlas for which the scale factor $c$ is consistent with unity but inconsistent with zero. The selection criterion is:
\begin{equation}
0 \leq (c - \delta c) \leq 1 \leq (c + \delta c),
\end{equation}
where $c$ is the best-fit scale factor and $\delta c$ is the associated error derived from the uncertainty in the least-squares fit. We find that 7 of the 265 signals in the atlas satisfy the above criterion, with varying levels of significance. 

We use the standard score, $\frac{c}{\delta c}$, to represent the significance. It measures the power of rejection of the null hypothesis; {\it i.e.}, the power with which the hypothesis that the scale factor is consistent with zero is rejected.  The standard score has units of standard deviation. We accept only those signals as being favored by the data in which the standard score is more than 2, which implies that these signals are consistent with the data and preferred by the data with greater than 2-standard-deviation confidence compared to the null hypothesis.  With this additional criterion, we find that 3 out of the aforementioned 7 signals are accepted and they have standard scores of 2.9, 3.3 and 3.4 respectively. The standard deviation of the residuals remains at $\sim 22.8$~mK in all the three cases. 

\begin{figure}[htbp]
\begin{center}
\includegraphics[scale=0.31]{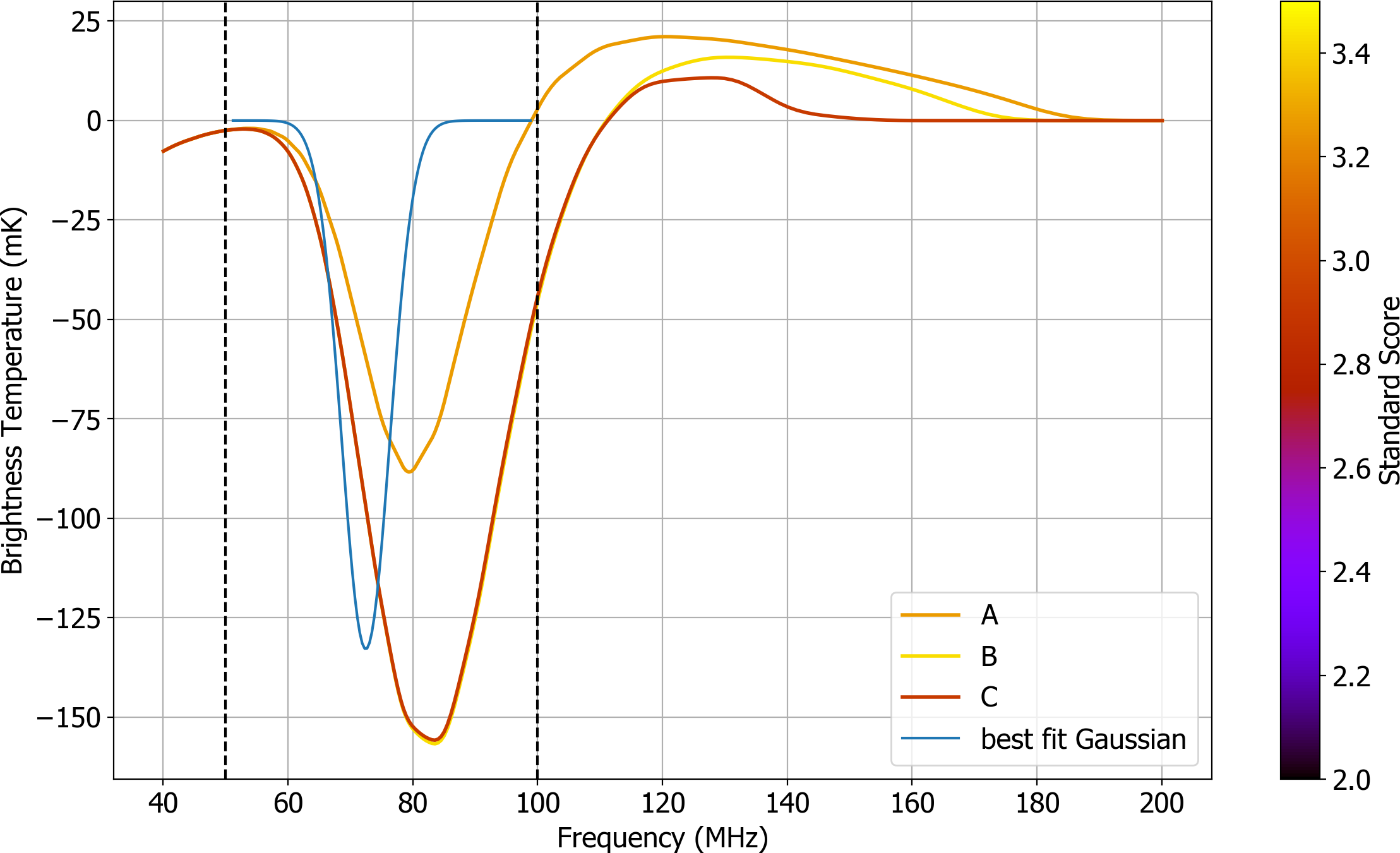}
\caption{The three 21-cm signal templates that are favored by the EDGES spectrum, of the atlas of 265 templates generated by \cite{2017MNRAS.472.1915C} as a sampling of the 21-cm signal space.  The signals are labeled $A$, $B$ and $C$ to link the traces to corresponding parameters in Table~\ref{tab:param}.  The favored templates are color coded by their standard score, which denotes the rejection of null hypothesis in units of standard deviations. Also overlayed for reference is the best-fit Gaussian. The vertical dashed lines show the frequency band of the analysis (50--100~MHz). }
\label{fig:7}    
\end{center}   
\end{figure}

\begin{figure}[htbp]
\begin{center}
\includegraphics[scale=0.29]{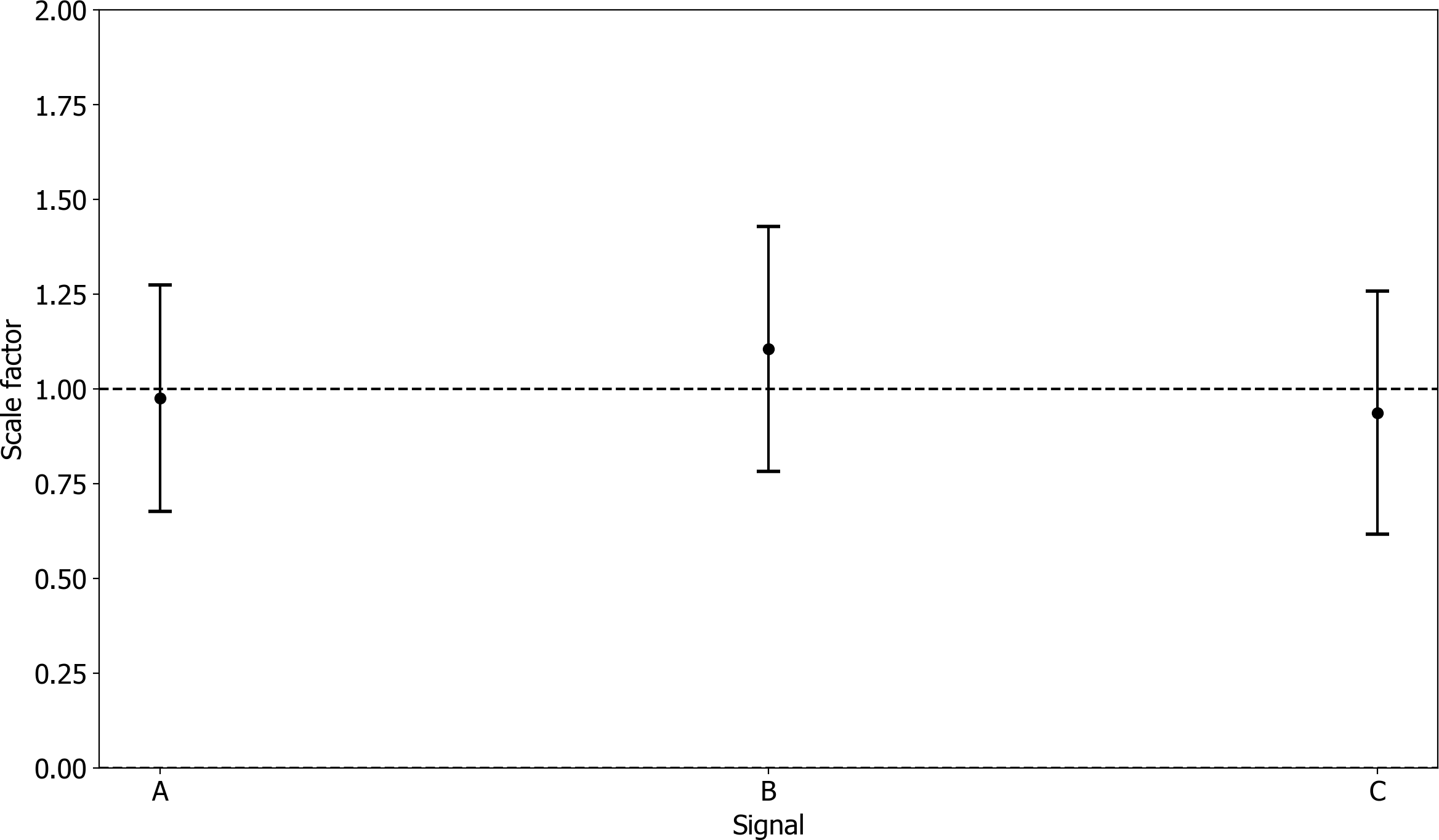}
\caption{The best-fit scale factors for the favored signals along with their associated errors.}
\label{fig:8}    
\end{center}   
\end{figure}

\begin{table}
{
\centering
\caption{Parameters of the 21-cm signals favored by data.}
\label{tab:param}
\begin{tabular}{||cccccc||}
\hline
$\textrm{Case}$ & $f_*$ & $V_c$ (km/s) & $f_X$ & $\tau$ & $R_{\rm mfp}$ (Mpc)  \\ 
\hline\hline
A&0.05 & 35.50 & 14.7 & 0.061 & 70  \\
\hline
B&0.158 & 35.50 & 1.58 & 0.066 & 70  \\
\hline
C&0.158 & 35.50 & 1.58 & 0.082 & 70  \\
\hline
\end{tabular}
}
\begin{tablenotes}
     \small
      \item The first column with heading `Case' has labels that denote signals shown in Fig.~\ref{fig:7}.  $f_*$ is the star formation efficiency, $V_c$ denotes the minimum virial circular velocity for star formation and $f_X$ represents the efficiency of the X-ray sources. $\tau$ is CMB optical depth and $R_{\rm mfp}$ represents the mean free path of ionizing photons. We refer the reader to \cite{2017MNRAS.472.1915C} for detailed description of these parameters.
    \end{tablenotes}
\end{table}

In Fig.~\ref{fig:7}, we show the three 21-cm templates that are favored by the EDGES data, out of the 265 atlas templates considered.  All these three yield best-fit scale factors close to unity in the joint fit of a maximally smooth function, sinusoid, and the template multiplied by the scale factor, which is a free parameter in the fit. Fig.~\ref{fig:8} shows the best-fit scale factors for the three signals selected by the data along with the extents of their errors. We list the astrophysical parameters for these signals in Table~\ref{tab:param} and discuss the allowed parameter space in Sec.~\ref{sec:dis}.

\section{Discussion}
\label{sec:dis}

The central frequency, the depth and the width of the absorption trough are regulated by high-redshift star and black hole formation. The low-frequency end of the absorption trough is related to the onset of star formation and the build-up of the Ly-$\alpha$ background radiation, while birth of the first X-ray sources triggers cosmic heating and marks the point of the deepest absorption. Relative timing of these events, as well as the rates of star formation and heating, determine the width and the depth of the absorption feature.   
Therefore, astrophysical properties such as star formation efficiency ($f_*$), minimum mass of the first star forming halos (M$_{\rm min}$ or, equivalently, circular velocity of these halos, $V_c$),  and properties of the first X-ray sources such as X-ray heating efficiency ($f_X$) and spectral energy distribution (SED)  can be constrained with the high-redshift 21-cm signal. Because the process of reionization is inefficient at high redshifts in the absence of massive galaxies, parameters such as the total CMB optical depth ($\tau$) and the mean free path of the ionizing photons ($R_{\rm mfp}$) do not play a significant role at high redshifts. For each high-redshift star formation scenario, the reionization parameters ($\tau$ and $R_{\rm mfp}$) can be tuned  to render the reionization history in agreement with available observations (e.g., \cite{2017MNRAS.465.4838G,2019MNRAS.484.5094G,2018arXiv180706209P}).  

We examined, above, 265 standard astrophysical models presented by \cite{2017MNRAS.472.1915C} and found that 3 of them are consistent with the EDGES low-band data. To produce this set of models, $\Lambda$CDM cosmology was assumed and the following astrophysical parameters were varied in the broadest range allowed by observations: $f_*$, $V_c$, $f_X$, $\tau$ and $R_{\rm mfp}$; in addition, hard versus soft X-ray SEDs were explored.  Earlier, 25 models out of this set, sharing low X-ray efficiency ($f_X = 0$ or $f_X = 0.1$) and rapid reionization, were found inconsistent with SARAS~2 data in the 110--200~MHz frequency range \cite[]{2017ApJ...845L..12S,0004-637X-858-1-54}. Extending the dataset of astrophysical models and varying, in addition to the above-mentioned parameters, the slope of the X-ray spectrum and the minimum frequency of X-ray photons,  a larger set of models was produced  and examined with the EDGES high-band data (Cohen et al. in prep. and \cite{2019arXiv190110943M}). Marginalizing over the rest of the parameters, \cite{2019arXiv190110943M} found that the High Band data alone disfavor low X-ray heating efficiency  ($f_X<0.0042$) and high minimum halo mass ($V_c>19.3$ km s$^{-1}$, equivalent to $\sim 1.3\times10^8$ M$_\odot$ at $z=10$). 

Location of the absorption feature within the EDGES low-band frequency range implies early onset of star formation and efficient X-ray heating, which is consistent with the non-detection of the 21-cm signal by SARAS~2 and EDGES high band.   Interpreting the EDGES detection, \cite{2019MNRAS.483.1980M} indicate that in order to ensure the absorption feature is within the band, high star formation efficiency of few percent  in faint galaxies below current detection threshold (halo masses of $10^8-10^{10}$ M$_\odot$) is required in addition to a luminous  X-ray population; while \cite{2019arXiv190103344S} found that inefficient star formation in minihalos is required. In agreement with these studies, we find that the 3 astrophysical models selected by our analysis above feature star formation efficiency  of  $f_* = 5\%$ (case A in Table~\ref{tab:param}) and $f_* = 15.6\%$ (cases B and C), minimum mass of star forming halos of $\sim 4\times10^8$ M$_\odot$ at $z=17$ (corresponding to $V_c = 35.5$ km s$^{-1}$) and moderately high  X-ray heating rate of $f_X = 14.7$ (case A) and $f_X = 1.58$ (cases B and C). Out of these three cases, $\tau$ of A and B are consistent with the latest {\it Planck} data (within 2$\sigma$ of the best-fit $\tau = 0.054\pm 0.007$ \cite[]{2018arXiv180706209P}), while $\tau$ of C is within 5$\sigma$. 

It may be noted here that the above exploration of the astrophysical parameter space is far from being exhaustive and a more rigorous analysis is required to establish the ranges of the astrophysical parameters favored by  the EDGES low-band data.  In particular, recognizing that the standard deviation of the residuals in the case of the 3 models selected are higher than that for a Gaussian-form absorption, and recognizing that the location of the peak absorption and also the full width at half maximum in the case of the 3 selected 21-cm profiles are somewhat different from those for the Gaussian-form signal, further exploration of the physical parameter space is needed towards generating models that might provide residuals with standard deviation closer to that for the case of a Gaussian absorption signal or, better still, closer to the expected measurement noise in the EDGES low-band data.

\section{Summary and conclusions}
\label{sec:4}

The EDGES claim of detection of redshifted 21~cm from Cosmic Dawn as a flattened absorption profile centered at 78~MHz, with amplitude of 0.5~K, has evoked a wide latitude of non-standard theoretical explanations for the signals' origin. However, it has also triggered discussions on the modeling of the data, and questioned whether the understanding of known systematics was adequate to account for, model and subtract their contributions.   Systematic errors might be caused by an inadequate foreground model, uncalibrated/unmodeled responses of the instrument to the sky, ground, or receiver noise, or a combination of these.  Accurate calibration, understanding and thereby modeling systematics, and analysis that separates systematics from any cosmological 21-cm signal, is the challenge towards making an unambiguous detection.  

We use the publicly available calibrated and time-averaged spectrum from EDGES low-band in the frequency range of 50--100~MHz.  The wide beams of radiometers such as EDGES average sky spectra over large angles and hence over a large number of sources with a distribution in radiation mechanisms and parameters; therefore, the average spectra cannot necessarily be fit with physical models with physically acceptable parameters.  Previous work \cite[]{2017ApJ...840...33S} suggests that such average spectra may be modeled using maximally smooth functions to the required accuracy. Therefore, we adopt the maximally smooth function as the representation for the foreground, in which the inherent constraints on higher order derivatives, and thereby the coefficients, makes it impossible to overfit, or fit out embedded high-order spectral structures.  Adopting such a foreground model, we are led to the conclusion that a flattened absorption signal in the data would have best-fit amplitude of 0.9~K, substantially deeper than the 0.5~K depth inferred by \citetalias{2018Natur.555...67B}.

Since the absorption depth inferred from our analysis, and also the depth inferred by \citetalias{2018Natur.555...67B}, are unexpected and require non-standard explanations, we have explored whether there may be uncalibrated systematics in the data that are the cause for the unexpected result.  The residuals to the fit with a maximally smooth foreground model indicates that there may be a sinusoidal-form systematic in the data; this has also been suggested previously by \citetalias{2018Natur.564E..32H}.  Sinusoidal residuals are indeed commonly seen in spectral radiometers and often arise from multipath propagation of wide-band noise within the signal path.  We find that the EDGES data are equally well fit with a foreground model plus sinusoidal systematic, compared to a foreground model plus flattened absorption.

Assuming that such a sinusoidal systematic does exist in the data, along with the foreground, we find that the data then favors the presence of an additional Gaussian absorption feature with amplitude $133$~mK: a depth that is well within the expectations from standard astrophysical cosmology.   Further, such a model is favored by information metrics over the model adopted by \citetalias{2018Natur.555...67B}, due to the $34\%$ lower variance in the residuals.  We then examined for whether the data might allow for any of the theoretically computed 21-cm profiles, using the atlas of templates provided by \cite{2017MNRAS.472.1915C} that are based on standard cosmology, constrained by previous observations, and represent a sparse sampling of the signal space.  We performed joint fits to the data using a maximally smooth function for the foreground and allowing for a sinusoidal systematic, and find that the data indeed favors a set of three templates with close to 3-standard-deviation significance standard scores.  Standard astrophysical models with the background $\Lambda$CDM cosmology, therefore, are found to fit the data well. The central frequency, the width and the depth of the best-fit absorption profile imply early onset of star formation in small dark mater halos of M$_{\rm h}\lesssim 10^9$ M$_\odot$ at few percent efficiency. In addition, moderately high X-ray heating is required with $f_X\gtrsim1$, which is consistent with standard assumptions for a high-redshift population of X-ray binaries.

We conclude that the EDGES data is indeed likely to have detected 21-cm from Cosmic Dawn; however, with an absorption depth that is consistent with standard models.

\section*{Acknowledgement}
We thank Anastasia Fialkov for useful discussions and sharing set of 21-cm signal templates that were used in exploring the consistency of the data with plausible models. We also thank Richard Hills for his comments and suggestions on an earlier version of this manuscript. We thank EDGES collaboration for making the data publicly available. SS acknowledges support from McGill Astrophysics postdoctoral fellowship.

\bibliographystyle{apj}

\bibliography{refer_ms}

\end{document}